\documentclass[aps,prl,superscriptaddress,showpacs,notitlepage,twocolumn,longbibliography]{revtex4-1}

\usepackage[english]{babel}
\usepackage{amsmath}
\usepackage{graphicx}
\usepackage{color}
\usepackage{xcolor}
\usepackage{braket}
\usepackage{verbatim}
\usepackage{amssymb}
\usepackage{amsmath}

\usepackage{soul}

\usepackage{subfigure} 
\usepackage{lipsum}%
\usepackage{hyperref}
\hypersetup{colorlinks=true}

\usepackage{graphicx}
\usepackage{dcolumn}
\usepackage{bm}

\usepackage{color}
\usepackage{braket}

\newcommand{\DFI}{Departamento de F\'isica, Facultad de Ciencias F\'isicas y Matem\'aticas, Universidad de Chile, Santiago 8370448, Chile}

\newcommand{\MIRO}{Millenium Institute for Research in Optics–MIRO, Universidad de Chile, Chile}

\newcommand{\ITMO}{School of Physics and Engineering, ITMO University, Saint  Petersburg 197101, Russia}

\begin{document}

\title{Long-range evanescent coupling through photonic molecules}

\author{Romina Abarca-Ram\'irez}
\affiliation{\DFI}
\affiliation{\MIRO}

\author{Diego Rom\'an-Cort\'es}
\affiliation{\DFI}
\affiliation{\MIRO}

\author{Maxim Mazanov}
\affiliation{\ITMO}

\author{Vlad~Simonyan}
\affiliation{\ITMO}

\author{Konstantin Rodionenko}
\affiliation{\ITMO}

\author{Maxim A. Gorlach}
\affiliation{\ITMO}

\author{Rodrigo A. Vicencio}
\email{rvicencio@uchile.cl}
\affiliation{\DFI}
\affiliation{\MIRO}

\begin{abstract}

Photonic molecules support the excitation of higher-order states, which are otherwise hard to access at individual waveguides. In this work, we demonstrate the resonant excitation of photonic molecular states which evanescently couple to single-mode waveguides. We implement the experiments on femtosecond laser written photonic structures and demonstrate an efficient resonant excitation of higher-orbital states, optimized at specific wavelengths and propagation distances. We suggest the use of long photonic molecules as long-distance photonic links, and demonstrate strong coupling for very distant waveguides separated by $127\ \mu$m. We apply this concept to a one-dimensional lattice and demonstrate the excitation of topological edge states emerging due to the third-order next-neighbour interactions. Our findings demonstrate effective long-range evanescent coupling which could be a concrete solution for fiber-based photonic chips, topological physics emerging from long-range interactions, or fundamental studies of initially uncoupled systems.  

\end{abstract}


\maketitle

The excitation of spatially extended stationary states opens the possibility for long-range (LR) transfer of energy. This is indeed the key of success in electronic transport in crystals~\cite{kittel_introduction_2005}, where a simple edge excitation allows perfect transmission of electronic waves through the entire system, leading to conducting materials. 

On the other hand, the hopping in solid-state insulating (tight-binding) systems or the evanescent coupling in photonics is based on localized and exponentially decaying wavefunctions at specific sites. This naturally sets a cutoff for the long-range interactions which decay exponentially with the distance~\cite{szameit_discrete_2005}. In addition, the tight-binding nature of the problem sets an upper speed limit on the propagation of excitations known as Lieb-Robinson bound~\cite{Lieb1972,Bravyi2006} or, more broadly, quantum speed limit~\cite{Deffner2017}. Artificial lattices of optical waveguides utilizing the evanescent coupling~\cite{lederer_discrete_2008} suffer from the same limitations, leaving multiple interesting topological and transport phenomena out of reach.

To remedy that, several numerical studies proposed distant coupling via the waveguiding
structure inserted below isolated waveguides~\cite{long2013} or via the overlapping photonic lattices~\cite{long2024a}. Very recently, these ideas started to translate into the experiments showing the long-range coupling between photonic moir\'e lattices~\cite{longrangeMoire}, quasi-BIC states in two-dimensional systems~\cite{long2024b,long2025}, in epsilon-near-zero materials~\cite{longrangeWang} and multilayered silicon photonic chips~\cite{Song2025}. In parallel, long-range coupling has been extensively explored in acoustic systems~\cite{longacoustic2017,RafaelMendez24}. However, the prospect of long-range coupling between the optical waveguides~-- the key components in photonic systems~-- remained elusive.

In this Letter, we resolve this challenge demonstrating evanescent coupling between the single-mode waveguides and long photonic structures~-- \textit{photonic molecules}~\cite{SPmolecules}. By tuning the geometric parameters of the structure and sweeping the excitation wavelength, we identify well-defined regions in the parameter space where the isolated waveguide excites spatially extended molecular states unlocking the long-range transfer of energy. We study the resonant excitation experimentally in the samples fabricated with the femtosecond (fs) laser writing technique and reveal a discrete set of resonances corresponding to the excitation of various molecular states. We harness this process as a \textit{long-range photonic link} and demonstrate the effective coupling of well separated waveguides, which otherwise can not interact. We suggest this concept as a practical solution for the coupling of optical fibers signals and their operation in photonic chips, which typically demands fan-in and fan-out methods. Finally, we use the LR coupling links in trivial 1D lattices and demonstrate topological edge localization, appearing due to the third-order next-neighbour coupling.

\begin{figure}[t!]
\centering
\includegraphics[width=1.\columnwidth]{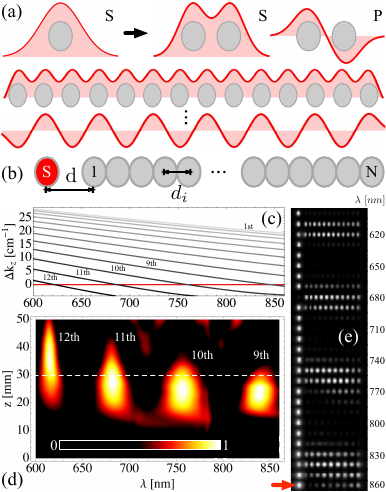}
   \caption{(a) Orbital states for different photonic configurations. (b) Excitation picture of a photonic molecule via a single mode waveguide. (c) Mismatch in the propagation constants for a molecule and single waveguide $\Delta k_z$ versus the excitation wavelength $\lambda$. The red line indicates $\Delta k_z=0$. (d) Normalized power at the molecule vs $z$ and $\lambda$, obtained from continuous modeling. (e) Intensity profiles at $z=30$ mm in the range $600-860$ nm, every $10$ nm.}  
\label{fig1}
\end{figure}

Higher-order photonic states are possible only for wider~\cite{SPmolecules} or stronger~\cite{guzman-silva_experimental_2021} waveguide structures, with a profile depending on the specific geometry. For example, elliptical waveguides, which are the standard in fs laser writing~\cite{szameit_discrete_2005}, are strongly anisotropic with a height approximately three times larger than the width. Therefore, the first higher-order modes emerging in such geometries are always vertically oriented~\cite{1Ddipole,MultiorbitalVic,Christina} and, following the atomic nomenclature, are simply called dipolar $P_y$ states~\cite{dipolarFBGraphene}. We have recently demonstrated the concept of photonic molecules~\cite{SPmolecules}, where we placed two single-mode elliptical waveguides very closely, such that they behave as a single wider entity. There, fundamental $S$ and horizontal $P_x$ higher states were excited and topological physics emerged on a Creutz-ladder geometry~\cite{li_2013}. Fig.~\ref{fig1}(a) schematically shows this concept, where a single-mode waveguide is used to form a small photonic molecule composed of two photonic atoms. In this case, two natural states emerge as a direct combination of wavefunctions: the bonding $S$ and the anti-bonding $P_x$ molecular modes~\cite{kaxiras_atomic_2003}. 

This method can be extended and used to create wider photonic structures, composed of a larger set of single-mode waveguides. Fig.~\ref{fig1}(a)-bottom shows a sketch of a long 1D photonic molecule composed of $14$ atoms. In this configuration, the molecular states are formed by the linear combinations of single-waveguide wavefunctions. As examples, we illustrate the fundamental in-phase and the higher out-of-phase molecular states.

Now, we explore the possibility of an indirect excitation of long molecular states via the evanescent coupling with a lateral single-mode waveguide [see sketch in Fig.~\ref{fig1}(b)]. We numerically compute the guided modes of an elliptical waveguide (red in figure) and the ones of a wider 1D structure formed by $N=31$ waveguides, with internal separation $d_i =4\ \mu$m and effective width of $120\ \mu$m (all in the wavelength range $\{600,860\}$ nm, and with both structures having the same refractive index contrast $\Delta n$~\cite{SM}). We evaluate the propagation constants $k_z$ and compare them in Fig.~\ref{fig1}(c), by defining $\Delta k_z\equiv k_{z}^{mol}-k_{z}^{S}$. We tune $\Delta n$ such that $\Delta k_z\approx 0$ around the 9th--12th molecular states, considering the experiments presented below.

We notice that a resonant evanescent mode interaction, from an excited S-mode at the isolated waveguide and a given state at the wider photonic structure, will be in general surrounded by two other molecular states. For example, Fig.~\ref{fig1}(c) shows that for $\lambda\sim680$ nm, a resonant coupling should occur among the single waveguide mode and the 11-th excited state, with the 10-th and the 12-th modes being around. That means that we could transit from one resonance into another by sweeping the waveguides parameters or by simply varying the excitation wavelength $\lambda$. This resonant process, however, also depends on the propagation coordinate $z$. For a perfectly tuned interaction~\cite{guzman-silva_experimental_2021}, the energy flows periodically between the waveguide and the photonic molecule, so that complete energy transfer occurs at $z = \pi (2 k + 1) / 2 C$, where $C$ is the coupling strength and $k$ is an integer.

For a deeper understanding of this resonant process, we study numerically~\cite{SM} the dynamics of the system sketched in Fig.~\ref{fig1}(b). We study the excitation of a single elliptical waveguide, separated by a distance $d=16\ \mu$m from a long 1D photonic molecule, formed by $N=31$ single-mode waveguides. We excite the isolated waveguide at $z=0$ and study the light dynamics along the propagation coordinate $z$ and  for different excitation wavelengths. Fig.~\ref{fig1}(d) shows the resulting dependence of the normalized power at the molecular structure, both on the propagation distance $z$ and the wavelength $\lambda$. We observe clear resonant molecular excitation at specific $\lambda$ and $z$-values, with sharper resonances for shorter wavelengths and longer distances. Fig.~\ref{fig1}(e) shows a set of intensity profiles obtained at $z=30$ mm and different $\lambda$, with the excited single waveguide located at the left (see red arrow). We observe a clear example of a resonant interaction between a single waveguide (photonic atom) and a photonic molecule possessing several photonic states. When their propagation constants match, there is a strong evanescent coupling from the \textit{S} state of the atom and the closest state of  the molecule. Of course, this resonant interaction is not perfectly sharp but, nevertheless, we observe a peaked distribution around a given specific state~\cite{guzman-silva_experimental_2021}. Hence, the system behaves as an effective photonic mode coupler~\cite{szameit_discrete_2005,lederer_discrete_2008}, with the energy flowing from one state into the other. Fig.~\ref{fig1}(e) clearly illustrates the quantized nature of this resonant interaction.

\begin{figure}[t!]
\centering
\includegraphics[width=1.\columnwidth]{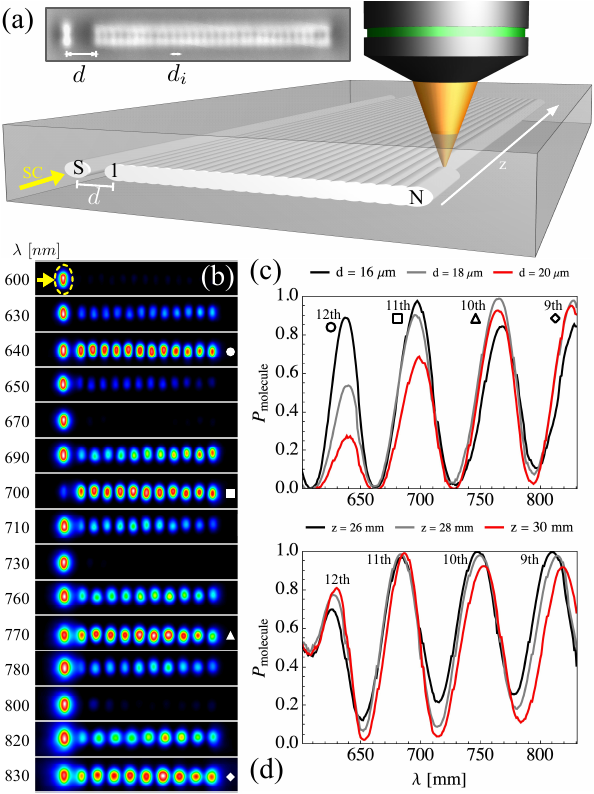}
   \caption{(a) Sketch for the fs laser writing technique. Inset: microscope image of a fabricated system. (b) Output intensity profiles for $d=16\ \mu$m and $z=30$ mm, at the indicated $\lambda$. (c) and (d) Extracted normalized power at the photonic molecule vs $\lambda$, for different distances $d$ (at $z=30$ mm) and for different propagation lengths $z$ (for $d=16\ \mu$m), respectively.}  
\label{fig2}
\end{figure}

We provide experimental evidence for the evanescent coupling between single-mode waveguides and elongated 1D photonic structures. We fabricate several photonic systems using the fs laser writing technique~\cite{szameit_discrete_2005} inside a borosilicate glass wafer as sketched in Fig.~\ref{fig2}(a). We study several molecular structures with different $N$-values, and also having different internal distances $d_i$~\cite{SM}. We select a photonic structure composed of $N=31$ individual waveguides with $d_i=4\ \mu$m, and a single-mode waveguide located at the left at a distance $d$ [see the inset of Fig.~\ref{fig2}(a)]. We characterize the system using a supercontinuum (SC) laser source, which allows us to ramp the excitation wavelength over a broad range $\lambda\in\{600,840\}$ nm. We tightly focus a polarized laser beam at the single-mode waveguide and continuously increase $\lambda$ to study the excitation of long molecular states at the wide photonic structure. Figure~\ref{fig2}(b) shows a set of different intensity output profiles for $d=16\ \mu$m and a propagation length of $z=30$ mm. At specific wavelengths, we observe the excitation of well-defined elongated molecular states as a result of an evanescent resonant coupling interaction. Remarkably, these very long excited states are homogeneously expressed along the~$\sim 120\ \mu$m-wide photonic structure, indicating minimal signs of spatial variation and therefore representing the ``true'' molecular modes. Overall, we find excellent agreement of the experimental results and the continuous numerical simulations presented in Fig.~\ref{fig1}.

Fig.~\ref{fig2}(b) shows the resonant and discrete nature of the evanescent coupling, very similar to the previous predictions considering individual waveguides and a continuous modification of the waveguide properties~\cite{guzman-silva_experimental_2021}. However, in the present case, the structure is fixed and the control external parameter is the excitation wavelength. By sweeping $\lambda$, we selectively tune the propagation constants of the isolated waveguide mode and the molecular states: $\Delta k_z\approx 0$, as Fig.~\ref{fig1}(c) suggests. For $\lambda=640$~nm we are able to distinctly excite the $12$-th mode, whereas at $\lambda=700$~nm the excitation of the $11$-th mode is almost perfect, with most of the energy being transferred to the molecule. For larger wavelengths $\lambda=770$~nm and $\lambda=830$~nm, we clearly excite the $10$-th and $9$-th modes. 

We deepen our experimental analysis by studying different distances $d$, and different propagation lengths $z$ for the molecular structure~\cite{SM}. Figs.~\ref{fig2}(c) and (d) show the extracted normalized power at the molecule versus $\lambda$, for the parameters indicated at the top of each panel. We observe clear resonant distributions at well-defined values of $\lambda$, with an optimal transfer at distinct optimal $z$. For example, Fig.~\ref{fig2}(c) shows that the excitation of the $11$-th mode is optimized for $d=16\ \mu$m and $z=30$ mm, with an almost perfect transfer $P_{{\rm molecule}}\sim 0.98$. For larger distances $d$ and smaller effective coupling constant $C$~\cite{Okamoto,norecidimer}, we notice that the optimal resonance occurs at larger wavelengths and with lower-order modes, due to the dispersive dependence of the coupling constants~\cite{diamondAPL23,2Drad25}. By fixing $d$, we notice in Fig.~\ref{fig2}(d) that the coupling distance also affects the maximum energy transport. For example, the $9$-th and $10$-th modes are perfectly excited for shorter propagation distances, while the $11$-th and $12$-th ones are optimized for larger ones, as their excitation occurs at shorter wavelengths and smaller $C$. All of these results are phenomenologically consistent with the simulations shown in Fig.~\ref{fig1}(d).

\begin{figure}[t!]
\centering
\includegraphics[width=1.\columnwidth]{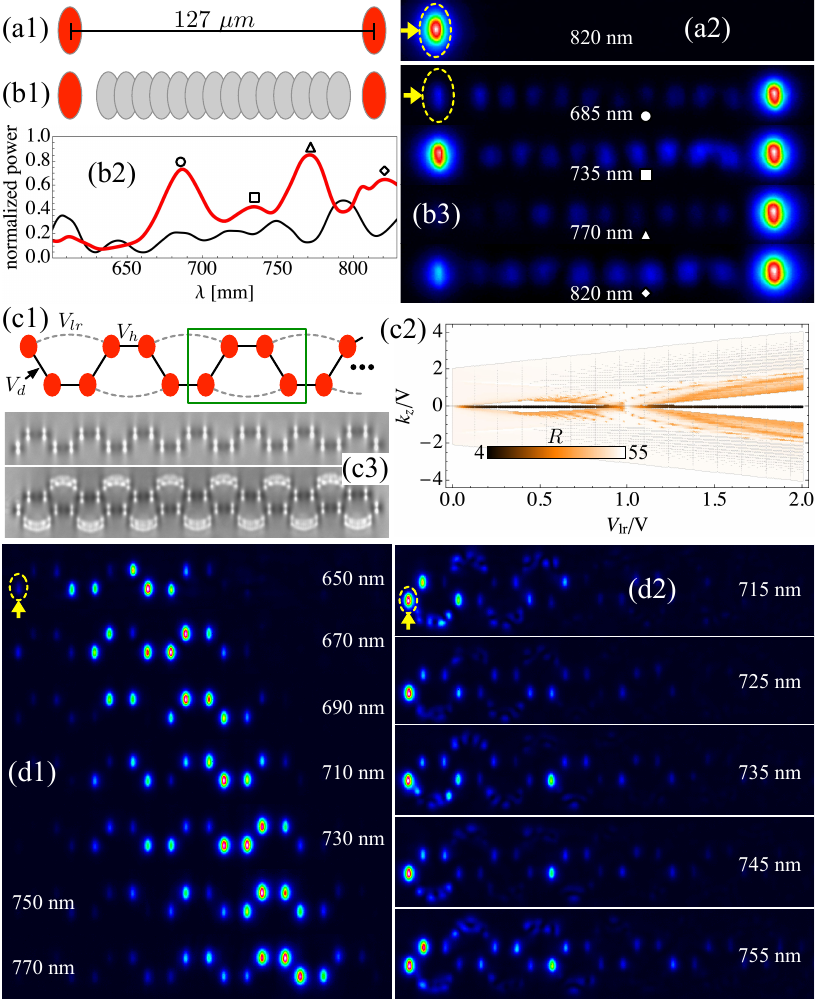}
   \caption{(a1) Sketch of two waveguides separated a distance $d=127\ \mu$m and (a2) output image at $\lambda=820$ nm. (b1) Sketch of a coupler including a photonic molecule as a link. (b2) Extracted normalized power vs $\lambda$ at the non-excited waveguide (red) and the molecule (black). (b3) Output images at the indicated wavelengths. (c1) A 1D lattice with horizontal $V_h$, diagonal $V_d$, and long-range $V_{lr}$ couplings. (c2) Linear spectra vs $V_{lr}/V$ (colors indicate the state's participation ratios). (c3) Microscope images of trivial and non-trivial fabricated lattices. (d1) and (d2) Output images, at the indicated wavelengths, after exciting the left edge site (see yellow ellipses) of a trivial and a non-trivial lattice, respectively.}  
\label{fig3}
\end{figure}

Next, we study possible applications of the observed multi-orbital coupling~\cite{MultiorbitalVic} using standard single-mode waveguides as input ports. One of the main technical problems when working with optical fibers and photonic chips is the insertion coupling~\cite{Thomson2015,crespi2013,szameitFiber}. A standard technique is based on $V$-groove configurations where bare fiber cores are spatially disposed at a minimal distance of $127\ \mu$m. Then, a fan-in mechanism is implemented to reduce the distance among the different waveguides, such that they can effectively interact inside the chip for classical or quantum operations~\cite{OBrien08}. A final fan-out scheme is afterwards necessary to take the processed information out of the chip again into new optical fibers; for example, for interferometry in astrophotonic applications~\cite{Thomson2024}. Both standard processes demand the fabrication of curved waveguides, which inherently introduces losses in the system~\cite{Okamoto}, affecting the operation at single-photon regimes. We start by fabricating only two waveguides at a distance of $127\ \mu$m [see Fig.~\ref{fig3}(a1)], both with a total glass length of $z=30$ mm. We inject light at the left waveguide and observe zero coupling for all $\lambda$ [see example in Fig.~\ref{fig3}(a2)]. Then, we insert a photonic molecule in between these two waveguides to generate a photonic link, as sketched in Fig.~\ref{fig3}(b1). We inject light at the left site and extract the normalized power at the opposite waveguide and molecule [see Fig.~\ref{fig3}(b2)]. We observe a resonant-like interaction which depends on $\lambda$, where the left waveguide first excites a photonic molecular state (black data), which then excites the waveguide at the right (red data). We observe a very good efficiency of this process, with $\sim 73\ \%$ of transfer at $\lambda=685$ nm, and a remarkable $\sim 85\ \%$ at $\lambda=770$ nm. The output profiles in Fig.~\ref{fig3}(b3) show some examples at specific wavelengths. In this case, the system behaves as an effective trimer and some light remains at the molecule, during the transition between the edges. The efficiency of this process could be optimized by designing different geometries of the photonic molecule, as well as by finely tuning the propagation distance for a specific $\lambda$ of interest. This transfer process resembles multimode interference devices~\cite{Soldano}, but having a transversal mode coupling instead of longitudinal.

Finally, we explore the use of long-range photonic links on a lattice. Recently~\cite{RafaelMendez24,Maffei2018Jan} it was suggested that the homogeneous 1D lattice could experience two topological transitions, after inserting a third-order coupling constant $V_{lr}$ [see sketch in Fig.~\ref{fig3}(c1)]. For $V_{lr}=0$, the lattice behaves trivially for $V_d=V_h=V$ as a dispersive single-band system~\cite{lederer_discrete_2008}. The addition of a long-range coupling transforms the system into a two-band non-trivial lattice supporting the edge states in the middle of the gap [see spectra in Fig.~\ref{fig3}(c2)]. The edge states are exponentially localized in space and have a small participation ratio $R$. Interestingly, this simple lattice allows two topological phases for $V_{lr}<V$ and for $V_{lr}>V$, with one or two edge states at each edge, respectively, distinguished by the winding number~\cite{Maffei2018Jan}.

We experimentally study this system by fabricating trivial and nontrivial configurations, the last including curved photonic links as shown in Fig.~\ref{fig3}(c3). Lattices were fabricated such that $V_d\approx V_h$ and $V_{lr}\sim 0.5 V_d$, over a broad range of wavelengths~\cite{SM}. We excite the lattices at the left edge site and summarize our results in Fig.~\ref{fig3}(d). 

For a trivial system Fig.~\ref{fig3}(d1), we observe a standard discrete diffraction pattern, with outer main lobes propagating away from the input position~\cite{lederer_discrete_2008}. Note that the SC excitation of lattices formed by the single-mode waveguides allows to map a wavelength-scan onto an effective increment of the propagation distance~\cite{diamondAPL23,2Drad25}. Hence Fig.~\ref{fig3}(d1) shows a typical ballistic transport expected for the dispersive systems. On the other hand, when the long-range coupling is switched on, the dynamics becomes completely different. Fig.~\ref{fig3}(d2) shows how an edge excitation mostly produces localization of the energy at the edge. Instead of traveling through the system, light remains well-trapped at the edge, showing a weak tail and a very weak dispersed background. Specifically, at $\lambda=725$ nm, the light keeps extremely localized at the edge and the photonic molecules contain almost no power~\cite{SM}, consistent with the picture where the molecules act as the third-order next-neighbour couplings. This phase corresponds to a large detuning $\Delta > V, V_{lr}$ between the modes in the main lattice and the molecular states, which is mapped onto the original model via the degenerate perturbation theory~\cite{Maffei2018Jan}. Interestingly, we find that the phases at other wavelengths, having smaller detuning $\Delta$ (including perfect degeneracy $\Delta = 0$), although having significant amplitudes at molecular modes, possess the same topological properties~\cite{SM}. These observations are a clear proof of the long-range coupling effect responsible for the emergence of the topological phase in an otherwise trivial 1D lattice.

In conclusion, we have demonstrated the use of long photonic molecular states as a powerful tool to couple distant waveguides, by matching the propagation constants for states having a very different spatial distribution. We numerically and experimentally demonstrated a resonant interaction between single-mode waveguides and spatially broader high-order photonic modes, occurring in well-defined parameter space regions. We use this strong excitation as a way of coupling very distant waveguides, which otherwise remain uncoupled. 

Moreover, we extended this concept towards a lattice geometry and applied long curved molecules to activate third-order next-neighbour coupling. This allowed us to observe a topological regime in an otherwise trivial 1D system, with a clear observation of edge localization as a concrete proof. Our observations could be extended to a broad range of devices and lattice geometries to study the physics emerging due to previously forbidden long-range photonic interactions. A possible route for further research could be the study of the interaction between the distant and initially isolated systems, as a way to extract quantum information~\cite{iso1,iso2,SatrianiBarra}.

\acknowledgements

This research was supported by Millennium Science Initiative Program ICN17$\_$012 and ANID FONDECYT Grant 1231313. Theoretical model for topological properties of 1D long-range-coupled array was supported by the Russian Science Foundation, grant No.~24-72-10069.


%

\end{document}